\RequirePackage{fix-cm}
\documentclass[smallextended]{svjour3}       % onecolumn (second format)
\smartqed  % flush right qed marks, e.g. at end of proof
\usepackage{graphicx}
\begin{document}

\title{On the survival of poor peasants%\thanks{Grants or other notes
}


\author{Andrea C. Levi     \and
        Ubaldo Garibaldi %etc.
}


\institute{A.C.Levi \at
              Department of Physics, University of Genoa, Italy \\
             % Fax: +123-45-678910\\
             % \email{fauthor@example.com}           %  \\
           \and
           U.Garibaldi \at
              Department of Physics, University of Genoa, Via Dodecaneso 33, 16146 Italy\\
              tel: 010-3536554 \\
		fax: 010-314218\\
\email{garibaldi@fisica.unige.it}}
\date{Received: date / Accepted: date}

\maketitle

\begin{abstract}
Previously, in underdeveloped countries, people tried to keep the prices
of food products artificially low, in order to help the poor to buy their
food. But it became soon clear that such system, although helpful for the
city poor, was disastrous for the peasants (who usually are even poorer), so
that hunger increased, instead of decreasing. More recently, thus, higher
prices have been imposed. But a high-price system does not solve the
problems. It helps, indeed, a peasant to buy in the city non-edible
products, but not to buy (more expensive) food products from other peasants.
The question is discussed here in more detail starting from the simplest
conceivable case of two peasants producing each a different food product
(bread and cheese, say), then generalizing to several food items and to any
number of peasants producing a given food item j. Like in every economic
system which wants to be sustainable, or able to reproduce itself in a
stationary state at least, prices are determined by the necessity of
exchanging \textquotedblleft means of production\textquotedblright\ among
\textquotedblleft industries\textquotedblright , except that here
\textquotedblleft industries\textquotedblright\ are replaced by working
peasants and \textquotedblleft means of production\textquotedblright\ are
replaced by food. It is found that prices must obey certain inequalities
related to the minimal amount of each food item necessary for survival.
Inequalities may be rewritten as equations and, in an important special
case, such equations give rise to a simple version of the matrix equation
used by famous authors to describe the economy.
%\subclass{MSC code1 \and MSC code2 \and more}
%\JEL{Jel code1 \and Jel code2 \and more}

\end{abstract}

\section{Introduction}
\label{intro}
Previously, in underdeveloped countries, the governments tried to keep food
prices artificially low, in order to help the poor to buy their food. But it
became soon clear that such system, although helpful for the city poor, was
disastrous for the peasants (who usually are even poorer), so that hunger
increased, rather than decreasing. More recently, thus, higher prices have
been imposed. But food prices non-uniformly higher, besides the possible
disadvantage of the city poor, do not solve the problems. They help, indeed,
a peasant to buy in the city non-edible products, but does not help him at
all to buy food products from some other peasants, if such products have
become more expensive.

In general exogenously fixed prices \footnote{%
Suppose e.g a government agency buying from producers and selling them what
they need at fixed prices} must satisfy some constraints to allow the
economy to stay alive. Further, if they are far from the \textquotedblleft
natural prices\textquotedblright\ (to be defined below)\cite{Sraf}, they
induce unfairness among individuals, and instability among sectors, possibly
to be suppressed by authority. The question will be discussed in more detail
starting with simple examples.

\section{Two peasants}

Let us suppose, to begin with, two food products to be needed in order to
survive, bread and cheese, say. Let 1 be the farmer producing the cereals
from which bread is made, 2 the shepherd raising the sheep from whose milk
cheese is made.

The monetary income of $1$, $R_{1}$, must be spent to buy cheese. The
inequality

\begin{equation}
R_{1}\geq p_{2}F_{2}  \label{uno}
\end{equation}

\noindent holds, where $p_{2}$ is the price of cheese and $F_{2}$ is the
minimal amount of cheese needed for survival.

On the other hand, how does $1$ obtain his income $R_{1}$? Selling bread. If
he produces the amount $Q_{1}$ of bread, he eats at least $F_{1}$ with his
family and sells $V_{1}$:

\begin{equation}
Q_{1}\geq F_{1}+V_{1}  \label{due}
\end{equation}

and, if $p_{1}$ is the price of bread, his income is

\begin{equation}
R_{1}=p_{1}V_{1}  \label{tre}
\end{equation}

Therefore

\begin{equation}
R_{1}\leq p_{1}(Q_{1}-F_{1})  \label{quattro}
\end{equation}

From (\ref{uno}) and (\ref{quattro}) the inequality

\begin{equation}
p_{2}F_{2}\leq p_{1}(Q_{1}-F_{1})  \label{cinque}
\end{equation}

obtains, i.e.

\begin{equation}
p_{2}/p_{1}\leq (Q_{1}-F_{1})/F_{2}  \label{sei}
\end{equation}

(survival of the farmer).

But a similar reasoning, considering the shepherd instead of the farmer,
yields the inequality

\begin{equation}
p_{1}/p_{2}\leq (Q_{2}-F_{2})/F_{1}  \label{sette}
\end{equation}

or%
\begin{equation}
p_{2}/p_{1}\geq F_{1}/(Q_{2}-F_{2})  \label{otto}
\end{equation}

(survival of the shepherd). (\ref{sei}) and (\ref{otto}) are compatible
(i.e. a price system exists under which both farmer and shepherd can live)
only if

\begin{equation}
F_{1}/(Q_{2}-F_{2})\leq (Q_{1}-F_{1})/F_{2}  \label{nove}
\end{equation}

which implies

\begin{equation}
F_{1}/Q_{1}+F_{2}/Q_{2}\leq 1.  \label{dieci}
\end{equation}

This seems to be an interesting conclusion; in fact, it is rather trivial.
Both $Q_{1}/F_{1}$ and $Q_{2}/F_{2}$ must always be larger than (or equal
to) $2$. If the size of the population is $2,$ 
\begin{equation}
Q_{i}\geq 2F_{i},i=1,2.  \label{10}
\end{equation}

A sizable production, thus, is required in each sector. E.g. if the farmer
produces the amount of bread needed for the survival of both, then the
shepherd must do the same for cheese. But if e.g. both have an
overproduction of $50\%$ only, there is no hope: with any price system, one
of the two succumbs.

The really interesting conclusions are (6) and (8), which may be written
together:

\begin{equation}
F_{1}/(Q_{2}-F_{2})\leq p_{2}/p_{1}\leq (Q_{1}-F_{1})/F2  \label{undici}
\end{equation}

yielding the admissible range of prices.

If e.g. $Q_{2}=2F_{2}$ and $Q_{1}=3F_{1}$, $\left( \ref{undici}\right) $
implies $1\leq p_{2}F_{2}/p_{1}F_{1}\leq 2.$In this interval both survive.
On the countrary if the price of the cheese dose falls below that of the
bread dose, the shepherd must draw on his savings (if and as long as
possible) to buy bread, or reduce the self-consumption $F_{2}$ or $F_{1}$;
both choices are unsustainable. $\ \ \ .$

Moreover, assuming the limiting value $p_{2}F_{2}/p_{1}F_{1}=1$, the
shepherd can just survive, while the farmer can save (or extra consume) a
dose of bread, or spend the saved value $p_{1}F_{1}$ elsewhere. An exogenous
price policy might produce deep inequalities in the population.

$\left( \ref{undici}\right) $ imply also

\begin{equation}
p_{1}F_{1}+p_{2}F_{2}\leq p_{1}Q_{1}  \label{dodici}
\end{equation}

(and similarly for $p_{2}Q_{2}$). The meaning of (\ref{dodici}) is
commonsensical: the final revenue has not to be lower than the costs of
production. It will become clearer below (see discussion after eq. (\ref{18}%
)).

\section{Many peasants, several sectors}

It is now easy to generalize to more than two food products (but the food
products necessary for survival are in all cases very few). It is also
convenient to deal with sectors rather than with single peasants: this
language shift is clearly allowed.

Let $N_{i}$ the number of peasants of the $i$-sector (i.e., producing the
food product $i$), while $N=\sum_{i}N_{i}$ will be the total number of
peasants of all sectors, the fraction of $i$-peasants being $n_{i}=N_{i}/N$.
Then (1) is replaced by

\begin{equation}
R_{i}\geq N_{i}\sum_{j\neq i}p_{j}F_{j}  \label{tredici}
\end{equation}

and (\ref{due}) and (\ref{quattro}) are replaced respectively by

\begin{equation}
Q_{i}\geq N_{i}F_{i}+V_{i}  \label{quattordici}
\end{equation}

and by

\begin{equation}
R_{i}\leq p_{i}(Q_{i}-N_{i}F_{i}).  \label{15}
\end{equation}

Taking together (\ref{tredici}) with (\ref{15})

\begin{equation}
\sum_{j\neq i}(p_{j}/p_{i})F_{j}\leq Q_{i}/(N_{i}-F_{i})  \label{16}
\end{equation}

i.e.

\begin{equation}
\sum_{j}(p_{j}/p_{i})F_{j}\leq Q_{i}/N_{i}  \label{17}
\end{equation}

or

\begin{equation}
N_{i}\sum_{j}p_{j}F_{j}\leq p_{i}Q_{i}.  \label{18}
\end{equation}

The inequality (\ref{18}) establishes an acceptable price system, since 
\begin{equation}
\sum_{j}p_{j}F_{j}=M  \label{18'}
\end{equation}%
denotes the cost of production \textit{per} worker. Thus (\ref{18}) says
that the cost of production of any sector cannot overcome its final revenue.
We can normalize the common factor of prices by posing e.g. $p_{i}=1$, or in
this special case (wherein the cost of production \textit{per} worker $M$ is
the same for all sectors) posing $M=1,$ so that all prices would be relative
to the "numeraire" arbitrary chosen. But we postpone this kind of choice,
and we mantain not normalized prices .

We have two basic inequalities for each $i$. First of all the "technical"
condition

\begin{equation}
Q_{i}\geq NF_{i}  \label{19}
\end{equation}%
must hold, implying that any food product $i$ is produced sufficiently for
everybody, assuming all people to have the same needs. (This equation
generalizes (\ref{10})). The second basic inequalities is, from (\ref{18}),(%
\ref{18'}), 
\begin{equation}
p_{i}\geq MN_{i}/Q_{i}  \label{19'}
\end{equation}%
$.$

\section{Equalities}

The inequalities (\ref{19}) and (\ref{18}) may be written as equalities.
Assuming that there is a surplus $Y_{i}=NF_{i}s_{i}\geq 0$ of food item $j$,
we rewrite (\ref{19}) as

\begin{equation}
Q_{i}=NF_{i}+Y_{i}=NF_{i}(1+s_{i})  \label{20}
\end{equation}

Similarly, we rewrite (\ref{18}) as

\begin{equation}
\frac{N_{j}}{Q_{j}}\left( \sum_{i}p_{i}F_{i}\right) \leq p_{j}.  \label{21}
\end{equation}

that is, posing 
\begin{equation}
A_{ij}=F_{i}\frac{N_{j}}{Q_{j}}\ \ \ ,  \label{22}
\end{equation}

$\sum_{i}p_{i}A_{ij}\leq p_{j}$ follows$,$and finally, introducing $%
r_{j}\geq 0,$

\begin{equation}
\sum_{i}p_{i}A_{ij}(1+r_{j})=p_{j}  \label{24'}
\end{equation}

Equation (\ref{24'}), except for the presence of different $r_{j}$'s, has a
form similar to equations well known in the economic literature\cite%
{Sraf,Pasin}. Typically, such equations describe the production of
industries of \ each sector using inputs from industries of other sectors;
in this case $r_{j}$ is the "rate of profit" of the $j$-sector. Although in
the present case of poor peasants the word "profit" sounds improper, or even
ironical, it is true that the well-being of the different sectors depends
precisely on the value of their $r_{j}$, and that, if for the $j$-sector $%
r_{j}$ happens to become negative, then the $j$-sector is destroyed, the
peasants of that sector can no longer survive, and the whole system is at
risk. In the case of exogenous prices, provided they are consistent with
inequalities( \ref{18}), the set of equations (\ref{24'}) can be solved for
the $r_{i}s$, which are different among sectors. Given that the prices are
exogenous, (\ref{24'}) can be solved for all $r_{i}$, ad the solution is 
\begin{equation}
r_{i}=\frac{p_{i}-\frac{N_{i}}{Q_{i}}M}{\frac{N_{i}}{Q_{i}}M},  \label{23'}
\end{equation}%
{}where $\frac{N_{i}}{Q_{i}}$ is the cost of production \textit{per }$j-$%
unit of food.

A general formula relates the rates of profit $r_{i}$ with the surplus
production realized in the different sectors:

\begin{equation}
\sum_{i}n_{i}\frac{1+r_{i}}{1+s_{i}}=1.  \label{24}
\end{equation}

Proof of (\ref{24}). Using (\ref{22}), eq. (\ref{24'}) may be written

\begin{equation}
\frac{N_{i}}{Q_{i}}(1+r_{i})\sum_{j}p_{j}F_{j}=p_{i}  \label{25}
\end{equation}%
But $\sum_{j}p_{j}F_{j}=M$, and $Q_{i}=NF_{i}(1+s_{i})$, hence (using $%
n_{i}=N_{i}/N$)

\begin{equation}
\frac{n_{i}}{F_{i}}\frac{1+r_{i}}{1+s_{i}}M=p_{i}.  \label{26}
\end{equation}%
Multiplying by $F_{i}$ and summing over $i$, we find again $%
\sum_{j}p_{j}F_{j}=M$ at the right-hand side, and (\ref{24}) is found.
Equation (\ref{24}), which can be rewritten usefully also as 
\begin{equation}
\sum_{i}\frac{N_{i}F_{i}}{Q_{i}}(1+r_{i})=1  \label{24"}
\end{equation}%
is a pure macroeconomical constraint, which connects physical magnitudes and
economical decisions. In fact, given (\ref{23'}), to fix exogenous rates of
profit is tantamount to fix exogenous prices. From (\ref{24}), given the
non-negativity of $r_{i}$ and the trivial $\sum_{i}n_{i}=1$, we see that in
the limiting case of all $s_{i}=0$ (that is a purely self-reproducible
system, no surplus) all $r_{i}$ are null, and then, \textit{via} (\ref{23'}%
), the sole possible prices are fixed, i.e. $p_{i}=\frac{N_{i}}{Q_{i}}.$ In
words, if the surplus vanishes, and the whole gross product is necessary to
the reproduction of the system, there is no room for any policy of exogenous
prices if the system has to stay alive. Note that only in the case of two
sectors (see (\ref{undici})) relative prices have both a lower and an upper
bound, fixed by the actual physical production. In the case of many sectors
the disequations (\ref{21}) and (\ref{19'}) furnish only a lower bound for
each price. The importance of (\ref{24}) or (\ref{24"}) is due to the
implicit double bound which prices have to respect for the system staying
alive.

A possible exogenous price fixing, suggested by the simple form (\ref{24}),
is to put $r_{i}=s_{i}.$ This position would reward sectors endowed by a
large surplus, and would punish less productive ones. This policy cannot but
be a temporary measure, as it would produce a bias for peasants abandoning
sectors worth of a larger production towards sectors which have no need to
increase\footnote{%
It is well known that an expanding economy the rate of growth is large as
the minimum surplus rate of its basic sectors. See\cite{Pasin}.
\par
.}.

For example, take $n_{1}=2/3,n_{2}=1/3,s_{1}=1/2,s_{2}=1/4$. Then, according
to Section $1$, the inequalities were $4/11\leq p_{2}F_{2}/(p_{1}F_{1})\leq
5/4,$ The prices induced by (\ref{24}) with $r_{i}=s_{i}$ are $%
p_{2}F_{2}/(p_{1}F_{1})=1/2..$

Another limit situation (see the next Section) is that of a policy that
would impose a uniform rates of profit for all sectors, but this would be
indistiguishable from a system left to itself and so "naturally" driven to
equalize different rates of profit. In this situation, with the above
parameters, $p2F2/(p1F1)=3/5$

\section{A single rate of profit}

Economic theory \cite{Sraf,Pasin,Lange},involves equations describing how
industries of a given sector use, for their production processes,
commodities obtained from industries of other sectors. In some cases such
equations have the general form

\begin{equation}
\sum_{i}p_{i}A_{ij}(1+r)=p_{j}.  \label{31}
\end{equation}%
They are a system of linear equations whose unknowns are the $n$ prices (the
number of sectors) and the (unique) rate of profit, and $A_{ij}$ are known
production coefficients. This is a homogeneous system, and \ prices are
determined up to an arbitrary common factor. Actually, the system is usually
more complicated than this, because it involves explicitly the competition
for the resources between capital and labour, which coincide in our simple
pure labour economy \cite{Pasi}. In our case of a simple pure labour
economy, the means of production are nothing else than the survival feeding
of workers.

Our equation (\ref{31}) seems to belong to this class, although the matrix $%
A_{ij}$ (see \ref{22}) is very simple, each entry being the product of $%
F_{i} $ (the $i-$dose of food) times $N_{j}/Q_{j},$ the labour for a unit of
product of the $j-$sector. Or more materialistically: $N_{j}F_{i}$ is the
quantity of $i-$commodity necessary to produce $Q_{j}.$ Given that the price
(of a commodity) is always implicitly referred to the unit (of that
commodity), $p_{i}A_{ij}=p_{i}F_{i}\frac{N_{j}}{Q_{j}}$ is the contribution
of the value of the $i-$commodity to the cost of a unity of the $j-$%
commodity.

\begin{eqnarray}
A_{ij} &=&F_{i}a_{j},  \label{32} \\
a_{j} &=&N_{j}/Q_{j},
\end{eqnarray}%
where $a_{j}$\ is the coefficient of labour of the $j-$sector, that is the
quantity of labour per $j-$unit (its reciprocal $h_{j}$\ is the $j-$%
productivity ). Note that the role of prices is quite different here: in (%
\ref{31}) they are endogenous, and can be obtained (up to a scale factor)
together with the (single) rate of profit $r$, while in (\ref{24'}) we had
different rates of profit $r_{j}$ for the different sectors. If prices are
fixed exogenously, and the rates of profit in the different sectors are
considerably different, the temptation for the individuals to pass to
sectors with high rate of profit is strong, but it is constrained by the
vinculum (\ref{20}). In words, sectors with low profit are compelled to be
populated, in order to keep alive the economic system; a sort of serfdom.
Instead if prices are endogenous and they are obtained by (\ref{31})
together with the single rate of profit, there is no reason for change of
sector, and the economy would be in a "natural" equilibrium.

Due to the simplicity of our matrix, (\ref{31}) can be solved exacty for the 
$n$ prices and the rate of profit $r,$ as the actual variables are only $n-1$
relative prices and $r$. Considering (\ref{31}) with $A_{ij}$ from ($\ \ \ref%
{32}\ )$ we have 
\begin{equation}
\frac{N_{j}}{Q_{j}}(\sum_{i}p_{i}F_{i})(1+r)=p_{j},j=1,..,n  \label{33'}
\end{equation}%
which implies 
\begin{equation}
p_{j}=V\frac{N_{j}}{Q_{j}}=Va_{j}  \label{34'}
\end{equation}%
If we put the constant $V=1,$ i.e 
\begin{equation}
p_{j}=\frac{N_{j}}{Q_{j}}  \label{34.2}
\end{equation}%
prices are measured\textit{\ }as the values produced by the work of the
peasants in a given time $t$\ (for example a day, a year). If $Q_{i}$\ is
the amount of food of type $j$\ produced in time $t$, and $\frac{N_{j}}{Q_{j}%
}$ is the amount of labour needed to produce a $j-$unit, then prices are
equal to "values"(in the sense of the classical school of economics).%
\footnote{%
From (\ref{20}), 
\begin{equation}
\frac{N_{i}F_{i}}{Q_{i}}=\frac{n_{i}}{1+s_{i}}
\end{equation}%
\par
It is interesting and correct (although at first sight unexpected) the fact
that the price of the $i$-dose is proportional to $n_{i}=N_{i}/N$: peasants
belonging to big sectors must obtain a relatively high price for food doses
produced by them. This is due to the fact that they must first of all feed
themselves and their families and, if they are many, the few doses left to
be sold must be exchanged with the many doses of the other commodities they
need to survive.}

Substituting (\ref{34.2}) in equations (\ref{33'}) we obtain\ 

\begin{equation}
\frac{1}{1+r}=\sum_{i}p_{i}F_{i}=\sum_{i}\frac{N_{i}}{Q_{i}}F_{i}
\label{35'}
\end{equation}

Note that, due to (\ref{19}), $\sum_{i}\frac{N_{i}}{Q_{i}}F_{i}\leq \sum_{i}%
\frac{N_{i}}{N}=1,$ and $r\geq 0$

\textit{\ }Hence

\begin{equation}
r=\frac{1}{\sum_{i}a_{i}Fi}-1.  \label{36'}
\end{equation}

Prices are proportional to $\frac{N_{j}}{Q_{j}}=a_{j}$, i.e the labour,
whereas up to here peasants appeared only \textit{via }the food they need to
survive. This food appears as capital to be anticipated, and the surplus is
shared among sectors proportionally to the capital. But in a pure labour
economy the capital is the "slave wage" only, which is proportional to the
number of workers. In the following we will assume endogenous ("natural")
prices and accordingly a single rate of profit $r$. A more formal approach
to "natural prices economy" is given in the following Chapter.

\section{Matrix approach}

In order to introduce the matrix approach, we turn to the mathematical form
of eq. (\ref{31}), which is attractive from a theoretical point of view. Eq.
(\ref{31}) is an eigenvalue equation, to which all the corresponding
theorems can be applied. It can be written in symbolic form as 
\begin{equation}
\mathbf{p\cdot A}(1+r)=\mathbf{p.}  \label{A1}
\end{equation}

where $\mathbf{p}=(p_{1},....,p_{n})$ is a row vector with as many
components as the number of commodities; $\mathbf{A}$ is a $n\times n$
matrix whose entries are the technical coefficients (quantity of $i$ needed
for a unit of $j$); and $r\geq 0$ is the rate of profit. In (\ref{A1}) $%
\mathbf{A}$ is given, so that we have $n+1$ unknowns. But due to
homogeneity, a price can be fixed at will, say $p_{1}=1$ (the numeraire, so
that all the other prices become relative), and the system has a unique
solution: the "natural price" $\mathbf{p}$, is the left eigenvector of $%
\mathbf{A}$ corresponding to the eigenvalue $\lambda =1/(1+r).$\textit{\ }%
Actually, by the Perron-Frobenius theorem (see Appendix of \cite{Pasin}), $%
\lambda $ is the maximum eigenvalue, it is not greater than $1$, and it is
the sole whose eigenvector eigenvector has all elements positive.\textit{\ }%
Further in our simple case the matrix factorizes, and it can be made even
simpler taking as units the individual survival doses of each commodity. In
this case we intoduce $X_{i}=Q_{i}/F_{i}$, whose the meaning is the number
of individual doses, and $A_{ij}=\frac{N_{j}}{X_{j}},$ the reciprocal of the
product number \textit{per capita }$\frac{X_{j}}{N_{j}}$ . Thus

$\mathbf{A=}\left( 
\begin{array}{cccc}
\frac{N_{1}}{X_{1}} & \frac{N_{2}}{X_{2}} & ... & \frac{N_{n}}{X_{n}} \\ 
\frac{N_{1}}{X_{1}} & \frac{N_{2}}{X_{2}} & ... & \frac{N_{n}}{X_{n}} \\ 
... & ... & ... & ... \\ 
\frac{N_{1}}{X_{1}} & \frac{N_{2}}{X_{2}} & ... & \frac{N_{n}}{X_{n}}%
\end{array}%
\right) $

\noindent\ all rows being equal\ (hence its determinant vanishes, but this
is irrelevant in the present treatment). \textit{\ }The maximum eigenvalue
is (the capital $R$ indicates the maximum value of $r$)

\begin{equation}
\lambda =\sum_{i}a_{i}=\sum_{i}N_{i}/X_{i}\leq 1  \label{A2}
\end{equation}

\noindent (see Sct. 5), where $a_{i}=N_{i}/X_{i}$\ is the coefficient of
labour of the $j-$sector, that is the quantity of labour per $j-$unit (its
reciprocal is a measure of the productivity ). The corresponding left
eigenvalue $\mathbf{p}$ is proportional to $\mathbf{a}=(a_{1},....,a_{n})$,
and the maximum profit rate is 
\begin{equation}
R=\frac{1}{\sum_{i}a_{i}}-1.  \label{A3}
\end{equation}

We have commented in Sct. 5 that the fact that the "natural price" is
proportional to the labour necessary to produce one unit (one dose, from now
on) looks strange in a scheme where all surplus is attributed to profit and
nothing to labour. In facts the complete form of (\ref{A1}) would be: 
\begin{equation}
\mathbf{p\cdot A}(1+r)+\mathbf{a}w=\mathbf{p.}  \label{A4'}
\end{equation}%
where $\mathbf{a}$ is the labour coefficient row vector, and $w$ is the
price of a unit of labour (the "human part" of labour, which exceeds pure
subsistence). Given that $\mathbf{p\cdot A}$ is the value of the means of
production, the value of the surplus $\mathbf{p-p\cdot A}$ is divided into a
capital part $\left( \mathbf{p\cdot A}\right) r$ and a labour part\ $\mathbf{%
a}w.$The opposite of (\ref{A1}) is%
\begin{equation}
\mathbf{p\cdot A}+\mathbf{a}w=\mathbf{p.}  \label{A5}
\end{equation}

\noindent where all surplus is allocated to labour. \ \ The formal solution
of (\ref{A5}) is 
\begin{equation}
\mathbf{p=a\cdot (I-A)}^{-1}w.\mathbf{.}  \label{A6}
\end{equation}

\noindent (\ref{A6}) says that prices are not simply proportional to $%
\mathbf{a,}$ but they are proportional to the vertically integrated labour
coefficients $\mathbf{a\cdot (I-A)}^{-1},$ a notion whose detailed
description we can now afford to omit \footnote{%
This notion is fundamental in many works of L.Pasinetti (see e.g.\cite{Pasi}%
, pag.75). The symbolic formulation of the system of quantities is $\mathbf{%
A\cdot Q+Y}=\mathbf{Q,}$where $\mathbf{Q}$ and $\mathbf{Y}$ are column
vectors representing the $n$ quantities and surpluses. Then $(\mathbf{I-}$ $%
\mathbf{A)\cdot Q=Y,}$ and $\mathbf{Q=}(\mathbf{I-}$ $\mathbf{A)}^{-1}%
\mathbf{\cdot Y.}$%
\par
\noindent The elements of the non-negative matrix $(\mathbf{I-}$ $\mathbf{A)}%
^{-1}$(which exists being $\mathbf{I-}$ $\mathbf{A}$ non singular) represent
the physical quantities $i$ nedeed, directly or indirectly, to obtain a unit
of $j-$commodity as surplus. This can be realized using the identity $(%
\mathbf{I-}$ $\mathbf{A)}^{-1}=\mathbf{I+}$ $\mathbf{A+A}^{2}+..+\mathbf{A}%
^{k}+...$, where each term adds the mean of production necessary to the
previous stage
\par
Then $\mathbf{a\cdot }(\mathbf{I-}$ $\mathbf{A)}^{-1}$ represents the
quantity of labour used, directly or indirectly, to obtain a physical unit
of any commodity.
\par
{}}. This because fortunately the two notions are in turn proportional if $%
\mathbf{a}$ is a letf eigenvector of $\mathbf{A}$, which is the basic
property of a pure labour economy\footnote{%
This remains true at any level of $r$ and $w$ in the general form (\ref{A4'}%
). In the whole treatise we suppose that sectors have no master, i.e. the
final product belongs to workers, who collectively anticipate the means of
production and share the net product. If so there is no difference is they
"think" themselves as capitalists or workers or any mixture of two. If there
were masters, though prices would not change, peasants' welfare would be
proportonial to $w$}; hence $\mathbf{p}=V\mathbf{a.}$ We can conclude that
in a pure labor economy prices are independent of the distribution of the
net product, and that sector shares of the surplus are the same, either
proportional to the invested "capital" or to labour.

\section{Exchanges with the city}

So far we have supposed that the countryside economy is a pure labour
economy, where there is no room for non-edible goods. If one wants to
introduce "luxury goods" as consumptions available to country people, one
has to introduce the sector to produce them. Coming back to the simple
example of Chapter 2, besides the farmer and the shepherd we introduce a
carpet seller, who, to produce $X_{3}$ carpets, needs bread, cheese and a
fraction of his product. The first observation is that the contryside can
have access to "luxury goods" only if the country surplus is enough to
substain one more worker, i.e. $X_{i}\geq 3$, see (\ref{10}). To avoid
inessential complications we consider the added value all given to profit
(we have treated the peasants as little interpreneurs, having used for them
the reduced form (\ref{A1}). The new system of linear equations is the
following, with $r=R$

\begin{equation}
\left\{ 
\begin{array}{c}
\begin{array}{c}
(p_{1}+p_{2})(1+R)=p_{1}X_{1} \\ 
\left( p_{1}+p_{2}\right) (1+R)=p_{2}X_{2}%
\end{array}
\\ 
\left( p_{1}+p_{2}+p_{3}\right) (1+R)=p_{3}X_{3}%
\end{array}%
\right.  \label{80}
\end{equation}

The different role of the third equation with respect to the previous two is
apparent. The first two equations are unchanged with respect to the old
closed system, they can be solved analytically, and the solution is just the
old one. If we put the system (\ref{80}) in matrix form,

\begin{equation}
\mathbf{A=}\left( 
\begin{array}{ccc}
\frac{1}{X_{1}} & \frac{1}{X_{2}} & \frac{1}{X_{3}} \\ 
\frac{1}{X_{1}} & \frac{1}{X_{2}} & \frac{1}{X_{3}} \\ 
0 & 0 & \frac{1}{X_{3}}%
\end{array}%
\right) ,  \label{81}
\end{equation}%
we see that the matrix is reducible, and the main submatrix (describing the
basic-commodities) is identical to the closed system. Posing $p_{1}=1$, i.e.
measuring values in terms of a dose of bread, we get $p_{2}=\frac{X_{1}}{X_{2}}.$ (See the note to the previous chapter). The rate of profit $R$ is 
\begin{equation}
R=\frac{1}{X_{1}^{-1}+X_{2}^{-1}}-1  \label{82'}
\end{equation}%
$,$ where $X_{1}^{-1}+X_{2}^{-1}$ is the maximum eigenvalue of the main
submatrix (see(\ref{A10})).$.$

All these variables being determined in the the main submatrix (describing
the basic-commodities), turning to the third equation we find 
\begin{equation}
p_{3}=\frac{X_{1}}{X_{3}-1-R},  \label{84}
\end{equation}%
and the system is completely solved. Note that (\ref{84}) must be taken with
care: we see that $X_{3}$, or rather $X_{3}-1$, must be large, i.e. the
production which exceeds its means, must be larger than $R$, to avoid
strange behaviour of $p_{3}$\cite{Sraf}

We can consider the effect on the countrymen of enlarging the economy to
comprehend the carpet seller. Neither the prices did change, nor the income
of each sector, that is $RM,$ i.e. the maximum rate of profit times the
value of the individual means of production. The sole advantage is given by
the possibility of owing carpets, besides bread and cheese, as commodities
to be consumed. For city workers there is the possibility of being sustained
by the country food, which they exchange with carpets.

Instead for the city sector also the price depends esplicitely on the rate
of profit, and this can be critical, due to the denominator of (\ref{84}).

Suppose that $X_{3}-1-R$ is small. The city sector is fragile, ruled by $%
\left( p_{1}+p_{2}+p_{3}\right) (1+R)=p_{3}X_{3},$ Its fragility is due to
the fact that, at odds with basic-commodities cases, $p_{1},p_{2}$ and $R$
are given exogenously from the point of view of the city. Otherways $p_{3}$
has no limit, as it does not affect any other price. For all basic equations 
$p_{i}X_{i}>p_{i}(1+R)$ holds by definition, because $R$ is endogenously
derived. But in case of a non-basic commodity, it may happen that $%
p_{3}X_{3}<p_{3}(1+R)\,,$\ being $R$ exogenous. If $X_{3}-1-R$ is small and
positive, i.e. the production is small, the price can grow without limit to
support the exchange with country commodities, to pay wages and retain a
"fair" profit. But in case of $p_{3}X_{3}<p_{3}(1+R)$, even an infinite
price is unable to satisfy the equation, as the total revenue is less than
the sole "fair" profit. \footnote{%
This example is discussed at some lemgth in \cite{Sraf}.}

\subsubsection{Productivity changes induced by exchanges with the city.}

The contact with the city is not only a chance for new types of
expenditures: it may induce the desire to improve one's technical ability,
or to better cultivation methods, like a farmer buying a plough. Starting
from equations (\ref{A4'}), we do not introduce new means of production: we
simply suppose that the farmer increases its productivity, while the
shepherd mantains its technical coefficients.

A first way to express the improved productivity is that of supposing the
farmer producing simply more bread, \textit{ceteris paribus.} Instead of (%
\ref{91}) we can write.%
\begin{equation}
\begin{array}{c}
(p_{1}+p_{2})(1+R)=\gamma p_{1}X_{1},1\leq \gamma \\ 
\left( p_{1}+p_{2}\right) (1+R)=p_{2}X_{2}.%
\end{array}
\label{94}
\end{equation}

The augmented production of bread changes the technical matrix, its
eigensystem and $R$, but in a simple way. $p_{2}=\frac{\gamma X_{1}}{X_{2}}$
grows linearly with $\gamma ,$ and the total income too, i.e. $\gamma
X_{1}-2+\frac{\gamma X_{1}}{X_{2}}(X_{2}-2)=2\gamma (X_{1}-\frac{X_{1}}{X_{2}%
})-2,$ and it is partitioned into two equal parts, since the means of
productions are equal in the two sectors. The technological improvement of
the farmer betters (economically at least) both sectors, and the increasing
income can support some other development.

In the second case we suppose that the farmer produces the same previous
amount working less (due to his increased productivity), i.e., using the
complete form derived from (\ref{A4'})

\begin{equation}
\begin{array}{c}
(p_{1}+p_{2})(1+r)+gw=p_{1}X_{1},0\leq g\leq 1 \\ 
\left( p_{1}+p_{2}\right) (1+r)+w=p_{2}X_{2}%
\end{array}
\label{91}
\end{equation}

The submatrix (\ref{81}) is unchanged, thus eigenvalues and eigenvectors do
not change, the maximum rate of profit is still (\ref{82'}). What changes is
the labour coefficient vector, which is not a left eigenvector of the
technical matrix unless $g=1.$ Hence we expect that natural prices depend on 
$r.$ But if we consider $r=R$ and $w=0$ the dependence on $g$ disappears,
and all economical values are the same. In this case the farmer can benefit
of the improved quality of life without any economical loss.

A third case of interaction with the city can be represented by the
enlargement of the two country sectors to comprehend a sector which produces
iron objects (like ploughs, spades,,,.), useful for improving the technique
of the farmer. At odds with luxury goods, let us suppose that the new
technique of the farmer needs a dose of iron, useless to the shepherd, while
the iron worker needs bread, cheese and one dose of iron. We suppose that
the production of bread increases, and that of cheese is the same, The
equations are: 
\begin{equation}
\left\{ 
\begin{array}{c}
\begin{array}{c}
(p_{1}+p_{2}+p_{3})(1+R)=p_{1}X_{1} \\ 
\left( p_{1}+p_{2}\right) (1+R)=p_{2}X_{2}%
\end{array}
\\ 
\left( p_{1}+p_{2}+p_{3}\right) (1+R)=p_{3}X_{3}%
\end{array}%
\right.  \label{92}
\end{equation}

Different from the case of luxury goods (\ref{80}), now the third price is
essential for the price of bread, so that the countryside is not independent
from the city sector. In this elementary case we can solve the system:

\begin{equation}
\left\{ 
\begin{array}{c}
p_{1}=1 \\ 
p_{3}=X_{1}/X_{3} \\ 
p_{2}=\frac{X_{1}(X_{3}-1-R)}{X_{2}X_{3}}%
\end{array}%
\right.
\end{equation}

The analytic solution for $R$ is omitted to avoid complications. The farmer
and the smith share the same part of the net product. If $\mathbf{X=}(8,3,2)$%
, i.e. the output of bread is doubled, that of cheese is invariant, and the
iron output is just enough for maintaining the economic flow, $R=.37$, and
the new prices are:

$p_{1}=1,p_{2}=\frac{8\ast (2-1-.37)}{3\ast 2}=\allowbreak 0.84,p_{3}=$ $%
\frac{8}{2}=\allowbreak 4.0$; the net product is $\mathbf{Y=}(5,0,0),$and
its value is $\mathbf{p.Y=}5.0.$

The means of productions of both bread and iron amount to $%
p_{1}+p_{2}+p_{3}=1+\allowbreak \allowbreak .84+4.0=\allowbreak
5.\,\allowbreak 84,$ and the income of the two sectors is $5.\,\allowbreak
84\ast .37=\allowbreak 2.\,\allowbreak 16.$ The income of the shepherd boils
down to $1.84\ast .37=\allowbreak 0.68,$ and the sum of the three incomes is
just $0.68+2\ast 2.16=\allowbreak 5.0=\mathbf{p.Y}$. The introduction of
iron objects has improved the income of the farmer (from $5/3=1.67$ to $%
2.\,\allowbreak 16$ doses of bread), while it has diminished the income of
the shepherd (from $5/3=1.67$ to $0.68$).

Unexpectedly, the fortune of the shepherd would rise abruptly if the smith
should produce one more dose. If $\mathbf{X=}(8,3,3),R=.64,\mathbf{Y=}%
(5,0,1),\mathbf{p.Y}=7.\,\allowbreak 67,$farmer's and smith's income becomes 
$3.\,\allowbreak 12$, while shepherd's one rises to $1.41$.

\end{document}